\documentclass[12pt,eqsecnum]{article}
\usepackage[dvips]{graphicx}
\usepackage{amssymb}
\usepackage{amsmath}
\usepackage{epstopdf}
\makeatletter

\@addtoreset{equation}{section}
\makeatletter

\newcommand{\Hb}{\frac{(-1)^m}{m ! }\left. \frac{\D^m W_{(I)}}{dZ^m}\right|_{Z=b}}

\newcommand{\D}{{\rm d}}

\newcommand{\dalm}{\kern1pt\vbox{\hrule height 0.9pt\hbox{\vrule width
0.9pt\hskip 2.5pt\vbox{\vskip 5.5pt}\hskip 3pt\vrule width 0.3pt}\hrule height
0.3pt}\kern1pt}

\def\b2hat{ {\hat b}_2 }

\def\HRZ{\eta(R,Z)}

\newcommand{\Ri}{{\cal R}}

\def\Ri{{\cal R}}

\newcommand{\ba}{\begin{array}}
\newcommand{\ea}{\end{array}}
\newcommand{\be}{\begin{equation}}
\newcommand{\ee}{\end{equation}}
\newcommand{\bea}{\begin{eqnarray}}
\newcommand{\eea}{\end{eqnarray}}

\newcommand{\Ialpha}{{\beta^{(I)}}}
\newcommand{\IW}{W_{(I)}}

\begin{document}

\begin{titlepage}
\vfill
\begin{flushright}
\today
\end{flushright}

\vfill
%\vskip 1.0cm
\begin{center}
\baselineskip=16pt
{\Large\bf 
Designer black holes from new 2D gravity\\
%A new class of 2D effective actions for non-singular spherically symmetric black holes\\
}
\vskip 0.5cm
{\large {\sl }}
\vskip 10.mm
{\bf Gabor Kunstatter${}^{a}$, Hideki Maeda${}^{b}$, and Tim Taves${}^{c}$} \\

\vskip 1cm
{
	${}^a$ Department of Physics, University of Winnipeg and Winnipeg Institute for Theoretical Physics, Winnipeg, Manitoba, Canada R3B 2E9\\
	${}^b$ Department of Electronics and Information Engineering, Hokkai-Gakuen University, Sapporo 062-8605, Japan\\
	${}^c$ Centro de Estudios Cient\'{\i}ficos (CECs), Casilla 1469, Valdivia, Chile. \\
	\texttt{g.kunstatter-at-uwinnipeg.ca, h-maeda-at-hgu.jp, timtaves-at-gmail.com}
     }
\vspace{6pt}
\today
\end{center}
\vskip 0.2in
\par
\begin{center}
{\bf Abstract}
\end{center}
\begin{quote}
    We present a family of extensions of spherically symmetric Einstein-Lanczos-Lovelock gravity. The field equations are second order and obey a generalized Birkhoff's theorem. The Hamiltonian constraint can be written in terms of a generalized Misner-Sharp mass function that determines the form of the vacuum solution. The action can be designed to yield interesting non-singular black-hole spacetimes as the unique vacuum solutions, including the Hayward black hole as well as a  completely new one.  The new theories therefore provide a consistent starting point for studying the formation and evaporation of non-singular black holes as a possible  resolution to the black hole information loss conundrum. 
\vskip 2.mm
\end{quote}
\end{titlepage}

%<<<<<<<<<<<<< PACS NUMBER >>>>>>>>>>>>>>>%
%\pacs{
%04.50.-h 	Higher-dimensional gravity and other theories of gravity
%04.50.Gh 	Higher-dimensional black holes, black strings, and related objects 
%04.60.-m 	Quantum gravity
%04.60.Ds 	Canonical quantization 
%04.60.Kz 	Lower dimensional models; minisuperspace models 
%} 

% CECS-PHY-13/09

%\maketitle

%\tableofcontents

\newpage

%======================================%
%<<<<<<<<<<<< SECTION I  >>>>>>>>>>>>>>%
%======================================%
\section{Introduction}

There exists considerable evidence for the existence of black holes in binary systems and at the center of most galaxies, including our own.
% In fact, the Event Horizon Telescope~\cite{Broderick2014} may soon be able to provide images of the event horizon of the black hole in our galaxy, a very exciting prospect. 
This observational evidence makes it imperative to resolve the theoretical puzzles surrounding black holes, most notable among them the so-called information loss conundrum. 
Interest in information loss has recently been vigorously renewed by the work of Almheiri, Marolf, Polchinski and Sully (AMPS)~\cite{Almheiri2013} 
%(See also~\cite{Susskind2012, Jacobson2013} 
who argued that the most conservative resolution to the conundrum is the existence near black-hole horizons of a firewall that breaks the quantum correlations between objects on the interior and those on the exterior. AMPS met with considerable resistance to their proposal due in large part to its implied violation of the strong equivalence principle near the horizon. Such quantum-gravity based violations are puzzling given that they must occur, for example, at the horizon of the recently observed~\cite{Wu2015} ten billion solar mass black hole where the curvature is a hundred orders of magnitude less than the Planck curvature.

 It has been known for some time~\cite{Frolov1979, Frolov1981, Markov1982, Roman1983} that the information loss problem could in principle be avoided if the singularity were replaced by a regular repulsive core. In this case the result of collapse and evaporation would be a complete spacetime containing a compact trapping horizon as opposed to an event horizon. This possibility was reiterated in a modern context in~\cite{Ashtekar2005} and put into concrete form by Hayward~\cite{Hayward2006}. More recently it has been actively discussed as an alternative to the firewall proposal~\cite{Hossenfelder2010, Bardeen2014, Hawking2014, Frolov2014}. 

There is substantial literature on static, non-singular black-hole spacetimes~\cite{Sakharov1966,Bardeen1968,Poisson1988,Dymnikova2005}. One particular metric that has been studied in some detail~\cite{Hayward2006, Frolov2014} is the Hayward metric, whose $n$-dimensional generalization is
\be
\D s_{(n)}^2 = - \left(1-\frac{l^{n-2}MR^2}{R^{n-1}+l^nM}\right)\D t^2 + \left(1-\frac{l^{n-2}MR^2}{R^{n-1}+l^nM}\right)^{-1}\D R^2+R^2\D\Omega_{(n-2)}^2, \label{Hayward-BH-n}
\ee
%matches (1.10) of long paper (LP) in the 4d case
{  where $\D \Omega_{(n-2)}^2$ is the line-element on the unit $(n-2)$-sphere and} we have defined a parameter $l$  that is proportional to the Planck length.  The Hayward metric has a non-singular de~Sitter core and curvature that is bounded above by $1/l^2$ for arbitrarily large mass\footnote{GK is grateful to Valeri Frolov for emphasizing the importance of this criterion.}.

In order to determine whether removing the singularity can solve the information loss problem, it is necessary to study quantitatively the dynamics of non-singular black-hole formation and evaporation. For example, the information must emerge from the black hole before the Bekenstein entropy bound~\cite{Bekenstein1994} is saturated, which would occur roughly in the so-called Page time~\cite{Page1993} when half the mass of the black hole has evaporated via the Hawking process.  Most investigations patch local regions to construct dynamical non-singular spacetimes that model the evaporation process~\cite{Hossenfelder2010, Bardeen2014, Hawking2014, Frolov2014}. More realistic models should be based on solutions to dynamical equations derived from a diffeomorphism invariant action that includes the back-reaction due to Hawking radiation. 
The essential features of this process likely reside in the spherically symmetric sector, so that one can hope to learn much by studying dimensionally reduced, effective two-dimensional (2D) actions. This was done to great effect in the 1990's in the context of 2D dilaton gravity~\cite{Grumiller2002} and has been revived more recently in~\cite{Ziprick2010, Ziprick2009, Taves2014}. 
%The dynamics of singular black-hole formation and evaporation have previously been studied analytically~\cite{Ashtekar2008} and numerically~\cite{Ayal1997, Ashtekar2011, Ashtekar2011a}. 
Until now, however, it has not been possible to use 2D gravity to construct realistic non-singular black holes such as (\ref{Hayward-BH-n}) with bounded curvature.
%Various models of non-singular black holes exist~\cite{Grumiller2003, Grumiller2004, Hayward2006, PK2008, Modesto2005, Modesto2010, PKZiprickThesis, Ziprick2010, Frolov2014}.

The purpose of this Letter is to present a new class of 2D gravity models that are obtained by extending spherically symmetric Einstein-Lanczos-Lovelock gravity in much the same way that 2D dilaton gravity extends general relativity. The resulting theories have all the desirable properties of 2D dilaton gravity, including the existence of a generalized Misner-Sharp mass function that renders the theories classically solvable. The corresponding one parameter family of solutions possess at least one Killing vector. The new actions describe a much larger class of spherically symmetric black-hole spacetimes than previously available, including those for which the maximum curvature is bounded by the Planck scale. They therefore provide a consistent starting point for studying the formation and evaporation of non-singular black holes as a possible alternative resolution to the black hole information conundrum.
{  In this Letter, we adopt units such that $c=\hbar=1$.}
%In the following we present the general class of actions, as well as the Hamiltonian analysis. Most importantly, we derive the integrability conditions on the Lagrangian functions for the existence of a  generalized mass function.  As a concrete example  we consider the action that yields the Hayward black hole as the unique solution. Details of the derivations, which are fairly long and involved, will be presented elsewhere.
%We start in Section II by reviewing briefly spherically symmetric EL and present the general class of extensions. We then do a complete Hamiltonian analysis and derive the generalized Misner-Sharp mass~\cite{Misner1964} for the theory. Section II then closes with the description of Einstein Gauss Bonnet theory as an illustrative example. Section III derives the most general solution, thereby proving the Birkhoff's theorem, and shows how to ``engineer'' lagrangians to obtain specific black hole solutions of interest.  In Section IV we show that there is a particularly interesting sub-class of theories, which we refer to quasi-Lovelock gravity, that are most closely related to the underlying EL theory. Finally we close with a summary a prospects for future work. Calculational details of the derivations are relegated to an appendix.

%======================================%
%<<<<<<<<<<<< SECTION I  >>>>>>>>>>>>>>%
%======================================%
\section{The action}
The action for the classes of theories we wish to consider is an extension of the spherically symmetric Einstein-Lanzcos-Lovelock (EL) gravity~{\cite{Lanczos1938, Lovelock1971}}.
The EL action in $n (\geq 4)$-dimensional vacuum spacetime is a sum of higher curvature terms.

Consider the spherically symmetric metric in $n\geq4$ dimensions:
\begin{align}
\D s_{(n)}^2 = {\bar g}_{AB}({y})\D {y}^A \D {y}^B + R^2({y}) \D \Omega_{(n-2)}^2, \label{eq:higherDg}
\end{align}
%matches (1.2) of LP
where ${\bar g}_{AB}({y})~(A,B=0,1)$ is the general two-dimensional Lorentzian metric and $R$ is the areal radius. It was shown in~\cite{Kunstatter2012, Kunstatter2013, Taves2013} that the dimensionally reduced spherically symmetric EL action can be written as
\bea
I_{\rm L}=\frac{{\cal A}_{(n-2)}}{16\pi G_{(n)}}\int \D^2{y}\sqrt{-{\bar g}}R^{n-2}\sum^{[n/2]}_{p=0}\alpha_{(p)} {\cal L}_{(p)},
\label{eq:SphericalLovelockAction}
\eea
%matches (2.8) and (2.12) of LP except the there it is I_{(2)}
where ${\cal A}_{(n-2)}$ denotes the area of $S^{n-2}$, {  $\alpha_{(p)}$ is the Lovelock coupling constant}, and
\begin{align}
\label{eq:LovelockSimplified}
 {\cal L}_{(p)} =& \frac{(n-2)!}{(n-2p)!} \Biggl[p\Ri[{\bar g}] R^{2-2p} 
+ (n-2p)(n-2p-1)\biggl\{\left(1-Z\right)^{p} +2pZ\biggl\}R^{-2p} \nonumber\\
&  + p(n-2p)R^{1-2p}\biggl\{1-(1-Z)^{p-1}\biggl\}(D_A R)\frac{(D^A Z)}{Z} \Biggr]. 
\end{align}
%matches (2.9) of LP
{  Here $\Ri[{\bar g}]$ and $D_A$ are the two-dimensional Ricci scalar and covariant derivative, respectively, and we have defined $Z$ by the norm squared of the gradient of $R$, namely}
\be
Z:=(D R)^2.
\ee
%matches (2.10) of LP

The generalized Misner-Sharp mass in Lovelock gravity was defined~\cite{Maeda2008,Maeda2011} as
\begin{align}
\label{qlm}
m_{\rm L}:=& \frac{(n-2){\cal A}_{(n-2)}}{16\pi G_{(n)}}\sum_{p=0}^{[n/2]}{\tilde
\alpha}_{(p)}R^{n-1-2p}[1-(DR)^2]^p,
\end{align}
where ${\tilde \alpha}_{(p)}:=(n-3)!\alpha_{(p)}/(n-1-2p)!$.
%matches (2.11) of LP
The gravitational field equations imply that $m_L=M$ is constant in vacuum, and yield the Schwarzschild-Tangherlini-type vacuum solution:
\begin{align}
\D s_{(n)}^2=&-f(R,M)\D t^2+f(R,M)^{-1}\D R^2+R^2\D \Omega_{(n-2)}^2. \label{f-vacuum}
\end{align}
%matches (2.13) of LP except M dependence
Since $(DR)^2=f(R,M)$ for the above metric, the form of $f(R,M)$ is determined algebraically from (\ref{qlm})~\cite{Wheeler1986}.

We propose the following natural extension of (\ref{eq:SphericalLovelockAction}) and (\ref{eq:LovelockSimplified}):
\bea
\label{eq:lagrangianL}
I_{\rm XL} 
&=&
\frac{1}{l^{n-2}}\int \D^2{ y} \sqrt{-{\bar g}}  \biggl\{ \phi(R)\Ri[{\bar g}]  +\HRZ+ \chi(R,Z) (D_A R)\frac{(D^A Z)}{Z}  \biggl\},
%\nonumber\\&=& I_{\rm G} +  I_{\rm L},
\eea
where $l^{n-2}:= 16\pi G_{(n)}/{\cal A}_{(n-2)}$.
%Matches (2.15) of LP but last term which was not divided by Z.
%where $\Phi(R)$, $\HRZ$ and $\chi(R,Z)$ are as yet arbitrary functions of $R$ and $Z=(DR)^2$. {\bf As is often done in 2D gravity, we have chosen to make the action dimensionless. $l$ is an arbitrary constant of dimension length. This constant derives ultimately from the higher dimensional Newton constant so we assume that is equal to the higher dimensional Planck length. Although it does not {\it a priori} need to be the same as the length that necessarily appears in the singularity resolved metric, it is simplest and most natural to assume as we do in the following that there is only one length scale, namely the higher dimensional Planck length. Consistency then demands that the dimensions of $\phi(R)$, $\HRZ$ and $\chi(R,Z)$ are $L^{n-2}$, $L^{n-4}$ and $L^{n-3}$, respectively.} 
Since the action (\ref{eq:lagrangianL}) lives only in two spacetime dimensions, $R$ and hence $Z:=(DR)^2$ are scalars. The metric (but not the action in general) can be lifted to $n$ dimensions by adding the angular piece $R^2d\Omega_{(n-2)}^2$ as in (\ref{eq:higherDg}), in which case $R$ and $Z$ recover their geometrical interpretations as areal radius and {  the norm squared of its gradient}, respectively. The novel feature of the action (\ref{eq:lagrangianL}) is that it contains arbitrarily high powers  of $(DR)^2$, and hence the ``velocity'' $R_{,t}$, {  where a comma denotes the partial derivative}. 
%We conjecture that this is the most general 2d Lagrangian involving only the metric and a scalar that yields second order equations for both.
 We will show in the next section that there exists a single integrability condition on the functions $\phi(R)$, $\HRZ$ and $\chi(R,Z)$ that guarantees the existence of a mass function in complete analogy with 2D dilaton gravity and EL gravity\footnote{We present only the results. Details will appear elsewhere~\cite{Kunstatter2015}.}. This leaves sufficient flexibility in the choice of remaining {  functions in the theory} to produce a large variety of interesting solutions.

%======================================%
%<<<<<<<<<<<< SECTION I  >>>>>>>>>>>>>>%
%======================================%
\section{Hamiltonian formalism and mass function}

The Hamiltonian analysis of spherically symmetric EL gravity has been extensively studied~\cite{Louko1997,Deser2005,Taves2012,Kunstatter2012,Kunstatter2013}. The following is based in large part on the methodology and results of~\cite{Kunstatter2012, Kunstatter2013} and uses the notation and conventions similar to~\cite{Taves2014}. We start with the general Arnowitt-Deser-Misner (ADM) metric in two spacetime dimensions:
\begin{equation}
  \label{eq:generalds2G}
  \D s^2 = -N^2\D t^2 + \Lambda^2(N_r\D t + \D x)^2.
\end{equation}
%matches notation of (3.1) of LP
In this parametrization we have\footnote{Note that the definition of $R_{,u}$ departs slightly from the notation in~\cite{Taves2014} where $y$ was used instead of $u$.}

\begin{align}
\label{eq:ZG}
&Z =-{R_{,u}}^2+b^2,
\end{align} 
where
\bea
b&:=&\Lambda^{-2}{R_{,x}}\label{eq:defb},\\
R_{,u}&:=&N^{-1}(R_{,t}-N_r R_{,x}).\label{eq:defRy}
\eea
%above three match (3.2, 3, 8) of LP

In order to proceed with the Hamiltonian analysis, it is useful to assume that $\chi(R,Z)$ has an expansion of the form
\be
\chi(R,Z)=\sum_I\Ialpha(R) \IW(Z)Z.
\label{eq:chi}
\ee
%This matches (2.18) of LP but I had to multiply by Z.
 In addition we assume that $\IW(Z)(=\IW(-{R_{,u}}^2 + b))$ has a Taylor expansion in $Z$ and hence in ${R_{,u}}^2$, so that
\be
\IW(Z) = \sum_{m} \Hb {R_{,u}}^{2m}.
\ee
%agrees with (3.9) and (3.10) of LP

A lengthy calculation~\cite{Kunstatter2015} yields the total Hamiltonian density to be
\bea
{\cal H}_{\rm T} %&=&  P_{\Lambda} \Lambda_{,t} + P_R R_{,t} - {\cal L}_{\rm TOT}\nonumber\\
   &=& N {\cal H} + N_r {\cal H}_r,
\label{eq:HT}
\eea
%agrees with (3.15) of LP except the subscript on H
with the Hamiltonian and diffeomorphism constraints given respectively by
\begin{align}
{\cal H}=& P_R R_{,u}   +\frac{1}{l^{n-2}}\Biggl[2\left(\frac{\phi_{,x}}{\Lambda}\right)_{,x}-{\Lambda} \HRZ   \nonumber\\
   & -2R_{,u}\sum_I\left( \Ialpha(R)\IW(Z)R_{,u} \frac{R_{,x}}{\Lambda} + 
      2\Ialpha(R) \frac{R_{,x}}{\Lambda} \sum_m \frac{\D}{\D b}\biggl(\Hb\biggl)\frac{{R_{,u}}^{2m+3}}{2m+3}
      \right)_{,x}\nonumber\\
  &  +2\Lambda R_{,u}\sum_I\Ialpha_{,R}  \sum_m \Hb \frac{{R_{,u}}^{2m+3}}{2m+3}
    - \sum_I\frac{R_{,x}}{\Lambda}\Ialpha(R){\IW(Z)}Z_{,x} \Bigg], \\
{\cal H}_r =& P_R R_{,x} - {P_\Lambda}_{,x}\Lambda,
\end{align}
%matches (3.17,18) of LP except := signs
where the $P_\Lambda$ and $P_R$ are the conjugate momenta to the fields indicated by their subscripts.  A crucial feature of the above is that $R_{,u}$ is a function of only $P_\Lambda$, $\Lambda$, and $R$, so that $P_R$ appears only in the first term of each constraint above. One can therefore implement the general procedure for obtaining the mass function by taking the linear combination of ${\cal H}$ and ${\cal H}_r$ that eliminates  $P_R$.  After some algebra, one obtains the remarkably simple expression:
\bea
\tilde{\cal H} 
  &:=&\frac{R_{,x}}{ \Lambda}{\cal H} -\frac{R_{,u}}{ \Lambda}{\cal H}_r \nonumber\\
 &=&\bigl(2\phi_{,RR}Z- \HRZ \bigl) R_{,x} + \bigl(\phi_{,R}-\chi(R,Z)\bigl) Z_{,x}.
\label{eq:IFinalH}
\eea
%matches (3.20,21) of LP

Equation~(\ref{eq:IFinalH}) is a key result of our paper. It guarantees the existence of a mass function ${\cal M}={\cal M}(R,Z)$ such that 
\bea
\tilde{\cal H}
   = - {\cal M}_{,x}.
\label{eq:dM}
\eea
%matches (3.25) of LP
providing that {  the functions in the action} satisfy the integrability condition:
\be
 \bigl(2\phi_{,RR}Z- \HRZ \bigl)_{,Z}  = \bigl(\phi_{,R}-\chi(R,Z)\bigl)_{,R}
\ee
%matches (3.26) of LP

In this case, the total Hamiltonian can be written
\be
H_{\rm T}  = \int \D x \left(-\tilde{N}{\cal M}_{,x} + \tilde{N}_r {\cal H}_r\right) +H_{\rm B},
\label{eq:H1}
\ee
%matches (3.22) of LP except subscript on H which is consistent with this paper
where the new Lagrange multipliers are
\bea
\tilde{N}:=   \frac{N \Lambda}{R_{,x}},   \qquad \tilde{N}_r :=  N_r + N\frac{R_{,u}}{R_{,x}}.  
\eea
%matches (3.23,24) of LP
Here $H_{\rm B}$ is the boundary term required to make the variational principle well defined.
{  Assuming asymptotic flatness, $\tilde{N}\to 1$ holds at infinity, while we have ${\cal M}=M=$constant for vacuum solutions}. Thus the required boundary term is just the ADM mass:
\be
H_{\rm B} = \int \D x\left(\tilde{N}{\cal M}\right)_{,x}\biggl|_{x=x_{\rm B}} = M,
\label{eq:HB}
\ee
%matches (3.33) of LP
 where $x=x_{\rm B}$ corresponds to the asymptotically flat region.
%Moreover $\tilde{N}_r\to 0$ in the asymptotically flat case so that this is in fact the only boundary term required.

We note that the mass function ${\cal M}$ commutes weakly with the constraints, and hence is a physical observable. It also commutes with the total Hamiltonian  and is therefore constant not only in space but time as well. 
%The above shows that in general requires one one to introduce the term containing $\chi(R,Z)$ as well as $\HRZ$. Otherwise one would need to introduce an integrating factor in order to define the mass function. This integrating factor would ruin the nice properties of the metric we are trying to achieve. 
%======================================%
\section{Designer black holes}

Given the existence of a mass function it is straightforward to derive the most general solution. One first needs to fix the diffeomorphism invariance. We choose  $R=x$ and $P_\Lambda=0$ to find a solution in the form of
\be
\D s^2 = -f(R,M) \D t^2+ f(R,M)^{-1}\D R^2.
\label{eq:XLmetric}
\ee
{  Because $Z=1/\Lambda^2$ in this gauge and ${\cal M}(R,Z)={\cal M}(R,\Lambda^{-2})$ is equal to a constant, $M$, for vacuum solutions, the metric function $f(={\Lambda^2})$ is obtained by solving an algebraic equation $M={\cal M}(R,f)$.}
%\bea
%{\cal M}(R,Z) &=& {\cal M}(R,1/\Lambda^2) = M\nonumber\\
%\to \qquad \Lambda^2 &=& f(R,M).
%\label{eq:Msoln}
%\eea
%above 3 agree with (4.6, 7) of LP 
The solution has at least one Killing vector, namely $\partial/\partial t$, and a single free parameter $M$. Up to possible degeneracies, the theory obeys a generalized Birkhoff theorem. 

%\section{Designer Black Hole Spacetimes}
We now have the machinery to ``design'' an action to produce any given static 2D metric of the form (\ref{eq:XLmetric}).  First invert the desired expression for $f=f(R,M)$
to write the mass $M$ in terms of $R$ and $f$.
{  The result $M={\cal M}(R,f)$ determines the covariant expression for the mass function ${\cal M}={\cal M}(R,Z)$.
%\be
%M=g(R,f).
%\ee
%This implies that the covariant expression for the mass function is
%\be
%{\cal M}  = g(R,Z).
%\ee
One then determines the functions in the action by calculating
\be
\frac{\partial {\cal M}}{\partial x} = \left.\frac{\partial {\cal M}}{\partial R}\right|_ZR_{,x}+
     \left.\frac{\partial {\cal M}}{\partial Z}\right|_RZ_{,x}
\ee
%above 3 match (4.8,9,10) of LP
and comparing to (\ref{eq:IFinalH}) to identify two conditions that determine $\HRZ$ and $\chi(R,Z)$ keeping $\phi(R)$ arbitrary. }

As a specific example, consider the Hayward metric (\ref{Hayward-BH-n}).
This leads to
\be
{\cal M}(R,Z) = \frac{(1-Z)R^{n-3}}{l^{n-2}}\left\{1-(1-Z)\frac{l^2}{R^2}\right\}^{-1},
\ee
%Agrees with (4.12) of LP except a factor of l^(n-2) was replaced by 2.  I left it.
{   from which we derive the functions in the action as}
\bea
\chi(R,Z)&=&  \phi_{,R} -\frac{R^{n-3}}{2\left\{1-l^2(1-Z)/R^2\right\}^2},\\
\HRZ&=& 2\phi_{,RR}Z - \frac{R^{n-4}}{\left\{1-l^2(1-Z)/R^2\right\}^2}
\left(1-\frac{n-5}{2}(1-Z)\frac{l^2}{R^2}\right).
\eea
%agree with (4.14) of LP (with l^(n-2) replaced by 2) but had to change the first sign in each
Note that $\chi(R,Z)$ does have the power series assumed in the derivation.

There is a particularly interesting sub-class of theories, which we call {\it designer Lovelock gravity}. It corresponds to the dimensionally reduced action (\ref{eq:SphericalLovelockAction}) for spherically symmetric EL, but with all the ${\tilde \alpha}_{(p)}$ potentially non-zero in $n$ dimensions. In this case the action can no longer be lifted to a higher dimension EL theory since the corresponding Lovelock terms vanish algebraically for $p>n/2$. However, this provides us with an interesting 2D generalization of the spherical theory that can be interpreted in one of two ways:\\
(i) the large coupling limit ${\alpha}_{(p)}=\infty$ for $p\ge [(n-1)/2]$, or\\
(ii) the large $n$ limit ($n\to \infty$).

In designer Lovelock gravity, the metric function $f(R)$ is  determined from a mass function that has an expansion of the form
\begin{align}
\frac{2M}{R^{n-1}}=& \sum_{p=0}^{\infty}{\tilde
\alpha}_{(p)}\biggl(\frac{1-f(R)}{R^2}\biggl)^p.
\label{eq:Qseries}
\end{align}
%This agrees with (4.15) but this time l^(n-2)/(n-2) is replaced by 2.  I suppose this doesn't matter in this section since this could get absorbed into the coupling constants.
Since the right-hand side is an infinite series it may be written as an analytic function of $(1-f)/R^2$ by choosing ${\tilde
\alpha}_{(p)}$ appropriately. 
 The Hayward black hole (\ref{Hayward-BH-n}) is clearly in this class.

The following non-singular, asymptotically flat black hole is also realized in designer Lovelock gravity: 
\begin{align}
f(R)=1+\frac{R^{n+1}}{2l^{n+2}M}\biggl(1-\sqrt{1+\frac{4l^{2n}M^2}{R^{2(n-1)}}}\biggl).
\end{align}
 The metric resembles the vacuum solution in Einstein-Gauss-Bonnet gravity~\cite{Wheeler1986,Boulware1985} but is realized  with only odd-order Lovelock terms in the action.

On the other hand, another frequently studied non-singular black hole, the Bardeen-type black hole,
\begin{align}
f(R)=1-\frac{l^{n-2}MR^2}{(R^2+M^{2/(n-1)}l^{2n/(n-1)})^{(n-1)/2}},
\end{align}
%This is a little off from (4.28) of the long paper.
cannot be expressed as an infinite series of the form (\ref{eq:Qseries}). In this case one must go to the full form of the action (\ref{eq:lagrangianL}).

%======================================%
%<<<<<<<<<<<< SECTION I  >>>>>>>>>>>>>>%
%======================================%
\section{Conclusions}

We have presented a new class of gravity theories in two spacetime dimensions that are readily understood as extensions of spherically symmetric Einstein-Lanczos-Lovelock gravity. They share many, if not all, of the latter's desirable properties: Second order equations and the existence of a mass function which in turn leads to a one parameter family of vacuum solutions with at least one Killing vector. In contrast to EL gravity, however, the extensions admit a large class of static non-singular black holes as vacuum solutions. By adding matter couplings and radiation back-reaction terms, one can study quantitatively the dynamics of the formation and evaporation of a larger class of interesting non-singular black holes than was previously available.

%Quantum effects are expected to resolve the classical singularity in general relativity. While the notion of space and time itself may very well break down in the vicinity of the classically singularity, it is possible, if not in fact likely, that the quantum effects could alter the conformal structure of the spacetime that is commonly thought to represent the formation and evaporation via Hawking radiation of black holes. The dynamics of this quantum corrected spacetime can in principle be modeled by an effective theory that admits non-singular black holes as solutions. In order to determine whether or not the absence of a singularity can solve the information loss conundrum, it is necessary to have such models so that the formation and evaporation of non-singular black holes can be studied quantitatively. The lagrangians presented here provide precisely such models. Another question of particular interest is whether or not any of the extended spherically symmetric theories that can in principle be lifted to covariant gravity theories in higher dimension. These issues are currently under investigation.

\subsection*{Acknowledgments}
%{\bf Acknowledgments}
GK is grateful to Jack Gegenberg, Viqar Husain, Jorma Louko, and Jon Ziprick for helpful conversations and comments on the manuscript.  
He also thanks Valeri Frolov for providing motivation to look for an action that could produce nice, non-singular, spherically symmetric black hole solutions. 
 HM thanks the Theoretical Physics group in University of Winnipeg for hospitality and support, where this work was started.
This work was funded in part by the Natural Sciences and Engineering Research Council of Canada.  Support was also provided by the Perimeter Institute for Theoretical Physics (funded by Industry Canada and the Province of Ontario Ministry of Research and Innovation).
This work has also been funded by the Fondecyt Grant No. 3140123. The Centro de Estudios Científicos (CECs) is funded by the Chilean Government through the Centers of Excellence Base Financing Program of Conicyt.

\bibliography{NonSingularPaper2015GK}
\bibliographystyle{unsrt} %unsrt %unsrturl /home/tim/Desktop/unsrturl/unsrturl(use this one)

%Zhang2015 and Kuchar1994 are not cited anywhere.

\end{document}